\documentclass[twocolumn]{el-author}

\usepackage{amsthm}
\usepackage{pmat}

\begin{document}

\title{Distributed MIMO coding scheme with low decoding complexity for future mobile TV broadcasting}

\author{M. Liu, M. H\'elard, M. Crussi\`ere and J.-F. H\'elard}

\abstract{A novel distributed space-time block code (STBC) for the next generation mobile TV broadcasting is proposed. The new code provides efficient performance within a wide range of power imbalance showing strong adaptivity to the single frequency network (SFN) broadcasting deployments. The new code outperforms existing STBCs with equivalent decoding complexity and approaches those with much higher complexities.}

\maketitle

\section{Introduction}
\begin{figure}[t]
\centering{\includegraphics[width=70mm]{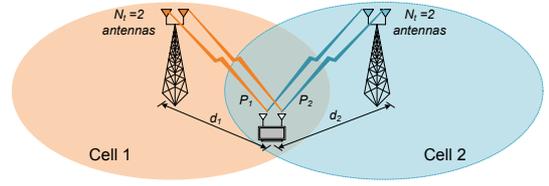}}
\caption{Distributed MIMO coding in a SFN broadcast scenario}
\label{fig:dis_mimo}
\end{figure}

The under-drafting Digital Video Broadcasting-Next Generation Handheld (DVB-NGH) standard firstly integrates the multiple-input multiple-output (MIMO) technique in the TV broadcasting aiming at efficiently delivering video services for the mobile terminals in the short future.
Current digital TV broadcasting systems widely adopt the single-frequency network (SFN), a network deployment where all transmitters simultaneously  send the same signal using same frequency band to enlarge the coverage without requiring extra frequency bands.
In order to keep compatibility with SFN, the future broadcasting systems should adopt MIMO coding schemes over adjacent broadcasting cells, namely distributed MIMO coding. Fig.~\ref{fig:dis_mimo} shows the distributed MIMO scenario considered in this work where MIMO coding is carried out not only among the two transmit antennas of the same cell (intra-cell coding) but also over two neighboring cells (inter-cell coding); the receiver equips two antennas forming a $4\times2$ MIMO transmission.
Due to the difference between the propagation distances ($d_1$ and $d_2$), the signals coming from different cells has different power, which affects the performance of the distributed MIMO coding.

Some space-time block codes (STBCs) such as 3D code, Jafarkhani code, L2 code, DjABBA code, Biglieri-Hong-Viterbo (BHV) code and Srinath-Rajan code which adapt to this scenario have been proposed in the literatures~\cite{nasser083d,hiltunen04four,srinath2009low} and references therein. However, all of them are either too complex for practical use or not robust to power imbalance of the received signal. In this work, we propose a novel STBC which is efficient, robust and simple for the distributed MIMO broadcasting scenario.

\section{A new proposal for $4\times2$ MIMO transmission}
All the aforementioned STBCs form codeword over four channel uses (i.e. $T=4$).
To reach full-rate, eight information symbols (i.e. $Q=8$) should be stacked in a codeword, which will result in high decoding complexity.
To reduce the complexity, we propose to encode four information symbols (i.e. $Q=4$) over two channel uses (i.e. $T=2$).
As proved later, this proposal guarantees full code rate while limiting the decoding complexity.
Recall that the Golden code~\cite{belfiore05golden} possesses high diversity with full rate in a $2\times2$ MIMO transmission.
Its encoding matrix is:
\begin{equation}\label{eq:Goldencode}
\textbf{C}_{\mathrm{GC}}=\begin{bmatrix}
        X_1(1) & X_1(2) \\
        X_2(1) & X_2(2) \\
        \end{bmatrix}
        =\frac{1}{\sqrt{5}}\begin{bmatrix}
        \alpha (s_1+\theta s_2) & \alpha (s_3+\theta s_4) \\
        i\bar{\alpha} (s_3+\bar{\theta} s_4) & \bar{\alpha} (s_1+\bar{\theta} s_2)\\
        \end{bmatrix}
        ,
\end{equation}
where $\theta=\frac{1+\sqrt{5}}{2}$, $\bar{\theta}=\frac{1-\sqrt{5}}{2}$, $\alpha=1+i\bar{\theta}$ and  $\bar{\alpha}=1+i\theta$ with $i=\sqrt{-1}$.
Stacking the four elements of the codeword in a vector $\mathbf{x}=[X_1(1),X_2(2),X_1(2),X_2(1)]^T$, the encoding can be further expressed by a linear transforming (denoted by a generator matrix $\mathbf{G}$) to information symbols $\mathbf{s}=[s_1,s_2,s_3,s_4]^T$:
\begin{equation}
\label{eq:Golden_encode}
\underbrace{\left [\begin{smallmatrix}
   X_1(1)\\
   X_2(2)\\
   X_1(2)\\
   X_2(1)\\
\end{smallmatrix}\right ]}_{\mathbf{X}}={\tiny
        \pmatset{1}{0.72pt}
        \pmatset{0}{2\pmatget{1}}
        \pmatset{2}{0pt}
        \pmatset{3}{0pt}
        \pmatset{5}{0pt}
\underbrace{\begin{pmat}[{.|.}]
\phi_1\cos\varphi & \phi_1\sin\varphi &    0   &   0  \cr
-\phi_2\sin\varphi & \phi_2\cos\varphi &     0   &   0\cr\-
0   &   0  & \phi_3\cos\varphi & \phi_3\sin\varphi \cr
0   &   0  & -\phi_4\sin\varphi & \phi_4\cos\varphi \cr
\end{pmat}}_{\mathbf{G}}}
{        \pmatset{1}{0.58pt}
        \pmatset{0}{2\pmatget{1}}
        \pmatset{2}{0pt}
        \pmatset{3}{0pt}
        \pmatset{5}{0pt}
\underbrace{\begin{pmat}[{}]
   s_1\cr
   s_2\cr
   s_3\cr
   s_4\cr
\end{pmat}}_{\mathbf{s}}},
\end{equation}
where $\sin\varphi=\frac{1}{\sqrt{1 + \bar{\theta}^2 }}$, $\cos\varphi=\frac{-\bar{\theta}}{\sqrt{1 + \bar{\theta}^2 }}$ with $\tan\varphi=-\frac{1}{\bar{\theta}}=\theta$ and
$\phi_1=e^{i\psi_{\alpha}}$,
$\phi_2=e^{i(\psi_{\bar{\alpha}}+\pi)}$,
$\phi_3=e^{i\psi_{\alpha}}$,
$\phi_4=e^{i(\psi_{\bar{\alpha}}+\pi)}$ with
$\psi_{\alpha}=\arctan(\bar{\theta})$ and
$\psi_{\bar{\alpha}}=\arctan(\theta)$.
The generator matrix $\mathbf{G}$ is a unitary, block diagonal matrix. Denote the two blocks as $\mathbf{G}_1$ and $\mathbf{G}_2$, respectively.

We know that Alamouti code is the most robust scheme in presence of imbalance of received signal power~\cite{nasser083d}.
With this knowledge and exploiting the property of Golden code, we propose a new STBC:
\begin{equation}\label{eq:3Dcodeshort}
\textbf{C}=\frac{1}{\sqrt{2}}\left [\begin{smallmatrix}
        X_1(1) & -X_1(2)^{\ast} \\
        X_2(1) & -X_2(2)^{\ast}\\
        X_1(2) & X_1(1)^{\ast} \\
        X_2(2) & X_2(1)^{\ast} \\
        \end{smallmatrix}
        \right]\!=\!\frac{1}{\sqrt{10}}\!\left [\begin{smallmatrix}
        \alpha (s_1+\theta s_2) & -\alpha^{\ast} (s_3^{\ast}+\theta^{\ast} s_4^{\ast}) \\
        i\bar{\alpha} (s_3+\bar{\theta} s_4) & -\bar{\alpha}^{\ast} (s_1^{\ast}+\bar{\theta}^{\ast} s_2^{\ast})\\
        \alpha (s_3+\theta s_4) & \alpha^{\ast} (s_1^{\ast}+\theta^{\ast} s_2^{\ast})\\
        \bar{\alpha} (s_1+\bar{\theta} s_2) & -i\bar{\alpha}^{\ast} (s_3^{\ast}+\bar{\theta}^{\ast} s_4^{\ast})\\
        \end{smallmatrix}
        \right].
\end{equation}
where $(\cdot)^{\ast}$ is the complex conjugate.
We note that, being different from the 3D code~\cite{nasser083d} which hierarchically encodes two Golden codewords with an Alamouti structure, the new code arranges one single Golden codeword in a similar manner for the sake of low decoding complexity while keeping the merit of robustness to power imbalance.
Since there are only $4$ information symbols in one codeword, the search space of the new code using ML decoder is $\mathcal{O}(M^4)$.
We will highlight some important properties of the proposed code and give concise proofs as follows.

\textit{\textbf{Property 1 (Full Rate):} The proposed STBC in (\ref{eq:3Dcodeshort}) achieves full rate.}

\textit{Proof:} The code rate of the proposed code is $R=Q/T=2$ which is equal to the number of received antennas $N_r$. Hence the new code is full rate. \hfill $\square$

\textit{\textbf{Property 2 (Rank of Codeword Difference Matrix):} The rank  of the codeword difference matrix $\mathbf{D}=\mathbf{C}-\mathbf{\hat{C}}$ is always equal to $2$ for all pairs of distinct codewords $\mathbf{C}$ and $\mathbf{\hat{C}}$, $\mathbf{C}\neq\mathbf{\hat{C}}$.}

\textit{Proof:} Suppose two distinct codewords $\mathbf{C}$ and $\mathbf{\hat{C}}$ are generated from two information symbol groups $\mathbf{s}=[s_1, s_2, s_3, s_4]^T$ and $\mathbf{\hat{s}}=[\hat{s}_1, \hat{s}_2, \hat{s}_3, \hat{s}_4]^T$ in which the corresponding symbols $s_j$ and $\hat{s}_j$ are not exactly the same, i.e. $s_j - \hat{s}_j$ being nonzero for at least one location $j$. At least one encoded symbol $X_m(n)$ is different from the corresponding $\hat{X}_m(n)$, $\forall m, n \in \{1, 2\}$. Hence, none of the two columns of matrix $\mathbf{D}$ is a zero vector.
With this fact and considering that the two columns of matrix $\mathbf{D}$ are orthogonal, these two columns are necessarily linearly independent.
Therefore, the column rank of the codeword difference matrix $\mathbf{D}$ is $2$. \hfill $\square$

Note that this is the maximum rank that can be expected from a $4\times2$ codeword matrix.

\textit{\textbf{Property 3 (Diversity Gain):} The diversity gain provided by the proposed code is $2N_r$.}

\textit{Proof:} The achievable diversity gain of a STBC is $rN_r$ given $N_r$ the number of receive antennas and $r$ the minimum rank of codeword difference matrix over all pairs of distinct codewords~\cite{tarokh98space}. With this fact and taking the result of \textit{Property 2}, it achieves \textit{Property 3}. \hfill $\square$

\textit{\textbf{Property 4 (Coding Gain):} The proposed code achieves a coding gain of $d_{\mathrm{min}}^2$, where $d_{\mathrm{min}}$ is the minimum Euclidean distance of the constellation adopted by the information symbols.}

\textit{Proof:} The coding gain~\cite{tarokh98space} provided by STBC code can be computed as $(\prod_{i=1}^{r} \lambda_i)^{\frac{1}{r}}$, where $\lambda_i$, $i=1,2,\ldots,r$, are the nonzero eigenvalues of the codeword distance matrix $\mathbf{E}\triangleq\mathbf{D}\mathbf{D}^{\mathcal{H}}$.
Taking into account the value of $\mathbf{D}$ with the proposed code, the two nonzero eigenvalue are $\lambda_1=\lambda_2=\sum_{m=1}^{2}\sum_{n=1}^{2}|X_m(n)-\hat{X}_m(n)|^2$ which is equal to the squared norm the first column of matrix $\mathbf{D}$ (denoted by $\mathbf{d}$).
Considering that the $\mathbf{X}$ is linearly coded from $[s_1, s_2, s_3, s_4]^T$, $\mathbf{d}$ can be expressed as $\mathbf{d}=\mathbf{G}(\mathbf{s}-\mathbf{\hat{s}})$.
Its squared norm is equal to $\|\mathbf{d}\|^2=(\mathbf{s}-\mathbf{\hat{s}})^{\mathcal{H}}\mathbf{G}^{\mathcal{H}}\mathbf{G}(\mathbf{s}-\mathbf{\hat{s}})=\|\mathbf{s}-\mathbf{\hat{s}}\|^2$ which is minimized when there is only one information symbol $s_j$ being different from the corresponding one $\hat{s}_j$ and these two symbols are the closest to each other among all combinations of constellation points. Therefore, the resulting minimum value of eigenvalues is $d_{\mathrm{min}}^2$ given $d_{\mathrm{min}}$ the minimum Euclidean distance of the constellation adopted by the information symbols. Taking into account the minimum rank of the proposed code $r=2$, the resulting coding gain is equal to $d_{\mathrm{min}}^2$. \hfill $\square$

\section{Low-complexity decoding algorithm}
Assuming that the channel remains constant over two time slots, the received signal $\mathbf{y}$ can be written as:
\begin{equation}\label{eq:rec_sig}
  \underbrace{\left [\begin{smallmatrix}
        Y_1(1)\\
        Y_2(1)\\
        Y_1^{\ast}(2)\\
        Y_2^{\ast}(2)\\
        \end{smallmatrix}
        \right]}_{\mathbf{y}}=\underbrace{{\tiny
        \pmatset{1}{0.72pt}
        \pmatset{0}{2\pmatget{1}}
        \pmatset{2}{0pt}
        \pmatset{3}{0pt}
        \pmatset{5}{0pt}
\begin{pmat}[{.|.}]
        h_{11} & h_{14} & h_{13} & h_{12} \cr
        h_{21} & h_{24} & h_{23} & h_{22} \cr
        h_{13}^{\ast} & -h_{12}^{\ast} & -h_{11}^{\ast} & h_{14}^{\ast} \cr
        h_{23}^{\ast} & -h_{22}^{\ast} & -h_{21}^{\ast} & h_{24}^{\ast} \cr
        \end{pmat}}}_{\mathbf{H}}  \underbrace{\left [\begin{smallmatrix}
        X_1(1)\\
        X_2(2)\\
        X_1^{}(2)\\
        X_2^{}(1)\\
        \end{smallmatrix}
        \right]}_{\mathbf{x}}+  \underbrace{\left [\begin{smallmatrix}
        W_1(1)\\
        W_2(1)\\
        W_1^{\ast}(2)\\
        W_2^{\ast}(2)\\
        \end{smallmatrix}
        \right]}_{\mathbf{w}},
\end{equation}
where $\mathbf{w}\sim \mathcal{CN}(0,\ 2\sigma^2\mathbf{I}_4)$ is complex white Gaussian noise vector and $\mathbf{H}$ is the channel matrix. The $(m,n)$th element of $\mathbf{H}$ is the channel gain from the $n$th transmit antenna to the $m$th receive antenna, modeled as independent and identically distributed (i.i.d) circular symmetric Gaussian distribution.
Dividing the channel matrix into two parts i.e. $\mathbf{H}=[\mathbf{H}_1\ \mathbf{H}_2]$ with $\mathbf{H}_1$ and $\mathbf{H}_2$ being the first and last two columns, respectively, and considering that the encoding matrix of the Golden code (\ref{eq:Golden_encode}) is block-wise diagonal, the received signal in (\ref{eq:rec_sig}) can be written in block form:
\begin{equation}\label{eq:rec_sig_blk}
\mathbf{y} = [\mathbf{H}_1\ \mathbf{H}_2] \begin{bmatrix}
        \mathbf{G}_1 & \mathbf{0}  \\
        \mathbf{0} & \mathbf{G}_2 \\
        \end{bmatrix}
        \begin{bmatrix}
        \mathbf{u} \\
        \mathbf{v} \\
        \end{bmatrix}
          + \mathbf{w}, = \mathbf{F}_1 \mathbf{u} +  \mathbf{F}_2\mathbf{v} + \mathbf{w},
\end{equation}
where $\mathbf{u}=[s_1,s_2]^T$, $\mathbf{v}=[s_3,s_4]^T$, $\mathbf{F}_j=\mathbf{H}_j  \mathbf{G}_j$ with $j\in\{1,2\}$.
The likelihood function of information symbols $\mathbf{u}$ and $\mathbf{v}$ provided $\mathbf{y}$ is~\cite{sirianunpiboon2010fast}:
\begin{align}
& p(\mathbf{y}|\mathbf{u},\ \mathbf{v})\propto \exp\big( -\frac{1}{2\sigma^2}\|\mathbf{y}-\mathbf{F}_1 \mathbf{u} -  \mathbf{F}_2\mathbf{v}\|^2 \big) \label{eq:likelihood_cond}\\
&= \exp\Big( -\frac{1}{2\sigma^2}(\mathbf{y}-\mathbf{F}_1 \mathbf{u})^{\mathcal{H}}(\mathbf{I}-\mathbf{F}_2(\mathbf{F}_2^{\mathcal{H}}\mathbf{F}_2)^{-1}\mathbf{F}_2^{\mathcal{H}})
    (\mathbf{y}-\mathbf{F}_1 \mathbf{u}) \Big) \nonumber \\
    &\times \exp\Big(-\frac{1}{2\sigma^2}\big(\mathbf{v}-\mathbf{\tilde{v}}(\mathbf{u})\big)^{\mathcal{H}}\mathbf{F}_2^{\mathcal{H}}\mathbf{F}_2\big( \mathbf{v}-\mathbf{\tilde{v}}(\mathbf{u})\big)\Big),
\end{align}
where
\begin{equation}\label{eq:ZF_s34}
\mathbf{\tilde{v}}(\mathbf{u})=(\mathbf{F}_2^{\mathcal{H}}\mathbf{F}_2)^{-1}\mathbf{F}_2^{\mathcal{H}}(\mathbf{y}-\mathbf{F}_1 \mathbf{u}).
\end{equation}
It is a zero-forcing solution of $\mathbf{v}$.
Hence, the likelihood function (\ref{eq:likelihood_cond}) can be solved by a conditional ML searching in two steps: 1) finding $\mathbf{\hat{v}}$ that maximizes  (\ref{eq:likelihood_cond}) with respect to a specific $\mathbf{u}$;
2) finding the best choice of $\mathbf{\hat{u}}$ and $\mathbf{\hat{v}}$ that maximizes (\ref{eq:likelihood_cond}) by
repeating the trial over all possibilities of $\mathbf{u}$. The optimal solution can be obtained as:
\begin{align}
    \mathbf{\hat{u}} &= \arg \min_{\mathbf{u}\in \mathbb{C}^2} \|\mathbf{y}-\mathbf{F}_1 \mathbf{u}- \mathbf{F}_2 \mathbf{\hat{v}}(\mathbf{u})\|^2,\\
    \mathbf{\hat{v}} &= \mathcal{Q}(\mathbf{\tilde{v}}(\mathbf{\hat{u}})),
\end{align}
where $\mathcal{Q}(\cdot)$ is the hard decision function and $\mathbf{\tilde{v}}(\mathbf{u})$ is given in (\ref{eq:ZF_s34}).
Since the estimation of $\mathbf{v}$ is achieved by linear processing, the proposed code is decoded through an exhaustive search over $M^2$ combinations of information symbols. Therefore, the overall computational complexity is of $\mathcal{O}(M^2)$. The exhaustive search can be replaced by the sphere decoder to further reduce the complexity~\cite{sirianunpiboon2010fast}.
This conditional ML searching offers a trade-off between the ML and linear decoding methods.

\begin{table}[tb]
\processtable{Worst case decoding complexity of STBCs}
{\begin{tabular}{|l|l|l|}\hline
\textbf{STBC scheme}& \textbf{Code rate}    & \textbf{Decoding complexity}\\\hline
Proposed code       & 2 &$\mathcal{O}(M^4)$ (ML) or $\mathcal{O}(M^2)$ (conditional ML)\\\hline
3D code             & 2 &$\mathcal{O}(M^8)$         \\\hline
BHV code            & 2 &$\mathcal{O}(M^6)$         \\\hline
DjABBA code         & 2 &$\mathcal{O}(M^7)$         \\\hline
Srinath-Rajan code  & 2 &$\mathcal{O}(M^{4.5})$     \\\hline
Jafarkhani code     & 1 &$\mathcal{O}(M^{2})$       \\\hline
L2 code             & 1 &$\mathcal{O}(M^{2})$       \\\hline
\end{tabular}}{\label{tbl_STs}}
\end{table}

\section{Comparison with the state-of-the-art STBCs}
We compare the proposed code with some state-of-the-art four-transmit-antenna STBCs, namely Jafarkhani code, DjABBA code, BHV code, Srinath-Rajan code (c.f.~\cite{srinath2009low} and references therein), L2 code~\cite{hiltunen04four} and 3D code~\cite{nasser083d}. The associated worst-case decoding complexities using sphere decoder are presented in Table~\ref{tbl_STs}.
The proposed code needs the lowest complexity $\mathcal{O}(M^4)$ among all rate-two codes to obtain the ML solution.
Note that the rate-one codes should use $M^2$-QAM to achieve the same throughput as the rate-two ones with $M$-QAM.
Therefore, the proposed code requires the same decoding complexity as rate-one codes to support the same system throughputs.
Moreover, it needs least complexity $\mathcal{O}(M^2)$ among all STBCs considered, if the conditional ML searching is employed.

\begin{figure}[h]
\centering{\includegraphics[width=70mm]{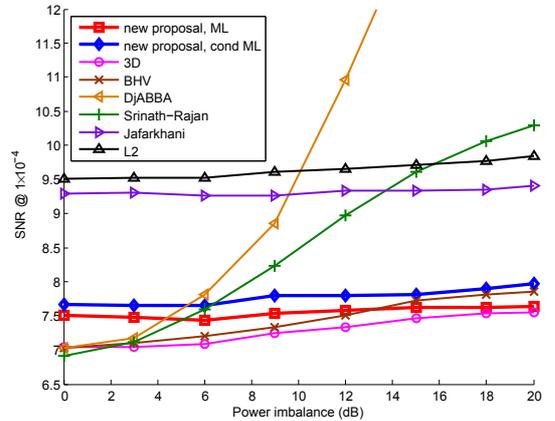}}
\caption{Performance comparison of the proposed code (thick lines) and some state-of-the-art codes (thin lines) in a distributed MIMO scenario.}
\source{DVB-NGH specifications: sampling frequency: $9.14$ MHz; FFT size: 4096; useful subcarrier: 3409; GI: $1/4$; QPSK for rate-2 codes, 16QAM for rate-one codes; channel coding: 16200-length LDPC, rate $4/9$; time interleaver size: $2.5\times10^{5}$ cells}
\source{DVB-NGH outdoor MIMO model with Doppler frequency of $33.3$ Hz}
\label{SNR_comparison}
\end{figure}

Fig.~\ref{SNR_comparison} shows the bit error rate (BER) performance of STBCs in different geographical locations using real DVB-NGH configurations.
The proposed code exhibits a strong resistance to the power imbalance.
It suffers less than $0.2$ dB performance loss in high power imbalance case using ML decoding.
In the contrary, most of the existing STBCs are not robust in distributed MIMO scenarios. Interestingly, when the power imbalance is high, namely greater than $15$ dB, the proposed code outperforms BHV, Srinath-Rajan and DjABBA codes and approaches 3D code which however requires much higher complexity.
The proposed conditional ML searching introduces negligible performance degradation with real system specifications while significantly reducing the complexity.

%\vfill\pagebreak

\section{Conclusion}
We have proposed a novel STBC for the next generation mobile broadcasting using distributed MIMO. The new code provides efficient performance with a large range of incoming signal power imbalance without requiring high decoding complexity.
It can be concluded that the new code achieves the best complexity-performance trade-off in the distributed MIMO broadcasting scenarios.

\vskip3pt
\ack{This work was supported by the European CELTIC project ``ENGINES''.}

\vskip5pt

\noindent M. Liu, M. H\'elard, M. Crussi\`ere and J.-F. H\'elard (\textit{Universit\'e Europ\'eenne de Bretagne (UEB),
INSA, IETR, UMR 6164, F-35708, Rennes, France})
\vskip3pt

\noindent E-mail: ming.liu@insa-rennes.fr

\end{document}